\documentclass[fleqn,usenatbib]{mnras}

\usepackage{newtxtext,newtxmath}
\usepackage[T1]{fontenc}

\DeclareRobustCommand{\VAN}[3]{#2}
\let\VANthebibliography\thebibliography
\def\thebibliography{\DeclareRobustCommand{\VAN}[3]{##3}\VANthebibliography}

\usepackage{graphicx}	% Including figure files
\usepackage{amsmath}	% Advanced maths commands
\title[The spectroscopic orbit of $\eta$ Car]{The orbital kinematics of $\eta$ Carinae over three periastra with a possible detection of the elusive secondary's motion}

\author[Strawn et al.]{Emily Strawn$^{1}$,
Noel D. Richardson$^{1}$,\thanks{E-mail: noel.richardson@erau.edu}
Anthony F. J. Moffat$^2$,
Nour Ibrahim$^{1,3}$,
\newauthor Alexis Lane$^{1}$,
Connor Pickett$^{1}$, 
Andr\'e-Nicolas Chen\'e$^4$,
Michael F. Corcoran$^{5,6}$,
\newauthor Augusto Damineli$^7$,
Theodore R. Gull$^{8,9}$,
D. John Hillier$^{10}$,
Patrick Morris$^{11}$,
\newauthor Herbert Pablo$^{12}$,
Joshua D. Thomas$^{13}$
Ian R. Stevens$^{14}$, 
Mairan Teodoro$^9$,
Gerd Weigelt$^{15}$\\
$^{1}$ Embry Riddle Aeronautical University, Department of Physics and Astronomy, 3700 Willow Creek Road, Prescott, AZ 86301, United States\\
$^2$ D\'epartement de physique, Universit\'e de Montr\'eal, Complexe des Sciences, 1375 Avenue Th\'er\`ese-Lavoie-Roux, Montr\'eal (Qc), H2V 0B3, Canada \\
$^{3}$ Department of Astronomy, University of Michigan, 1085 S. University, Ann Arbor, MI 48109, USA\\
$^4$ NSF’s NOIRLab, 670 N. A’ohoku Place, Hilo, Hawai’i, 96720, USA \\
$^5$ CRESST \& X-ray Astrophysics Laboratory, NASA/Goddard Space Flight Center, Greenbelt, MD 20771, USA \\
$^6$ The Catholic University of America, 620 Michigan Ave., N.E. Washington, DC 20064, USA \\
$^7$ Universidade de S\~ao Paulo, Instituto de Astronomia, Geof\'isica e Ci\^encias Atmosf\'ericas, Rua do Mat\~ao 1226, Cidade Universit\'aria, S\~ao Paulo, Brasil \\
$^8$ Exoplanets \& Stellar Astrophysics Laboratory, NASA/Goddard Space Flight Center, Greenbelt, MD 20771, USA\\
$^9$ Space Telescope Science Institute, 3700 San Martin Drive. Baltimore, MD 21218, USA \\
$^{10}$ Department of Physics \& Astronomy \& Pittsburgh Particle Physics, Astrophysics, \& Cosmology Center (PITT PACC), \\
University of Pittsburgh, 3941 O’Hara Street, Pittsburgh, PA 15260, USA \\
$^{11}$ California Institute of Technology, IPAC, M/C 100-22, Pasadena, CA 91125, USA \\
$^{12}$ American Association of Variable Star Observers, 49 Bay State Road, Cambridge, MA 02138, USA \\
$^{13}$ Department of Physics, Clarkson University, 8 Clarkson Ave, Potsdam, NY 13699, USA \\
$^{14}$ School of Physics and Astronomy, University of Birmingham, Birmingham B15 2TT, UK\\
$^{15}$ Max Planck Institute for Radio Astronomy, Auf dem H\"ugel 69, 53121 Bonn, Germany
}

\date{Accepted XXX. Received YYY; in original form ZZZ}

\pubyear{2022}

\begin{document}
\label{firstpage}
\pagerange{\pageref{firstpage}--\pageref{lastpage}}
\maketitle

\begin{abstract}
%\textbf{(just a place holder)}
The binary $\eta$ Carinae is the closest example of a very massive star, which may have formed through a merger during its Great Eruption in the mid-nineteenth century. We aimed to confirm and improve the kinematics using a spectroscopic data set taken with the CTIO 1.5 m telescope over the time period of 2008--2020, covering three periastron passages of the highly eccentric orbit. We measure line variability of H$\alpha$ and H$\beta$, where the radial velocity and orbital kinematics of the primary star were measured from the H$\beta$ emission line using a bisector method. At phases away from periastron, we observed the He~{\sc{ii}} 4686 emission moving opposite the primary star, consistent with a possible Wolf-Rayet companion, {although with a seemingly narrow emission line}. This could represent the first detection of emission from the companion.

\end{abstract}


%%%%%%%%%%%%%%%%%%%%%%%%%%%%%%%%%%%%%%%%%%%%%%%%%%

%%%%%%%%%%%%%%%%% BODY OF PAPER %%%%%%%%%%%%%%%%%%

\section{Introduction}

%Massive stars

The binary star system $\eta$ Carinae is known for being one of the most massive and luminous binaries in our local galaxy \citep{2012ASSL..384.....D}. The two stars are locked in a highly eccentric orbit \citep{Damineli1996, 1997NewA....2..107D}. Enveloping these stars is the Homunculus nebula which was formed by a large eruption in the mid-nineteenth century \citep[e.g.,][]{Currie1996}. The Great Eruption that formed the Homunculus nebula was recently modeled to be the product of a binary merger in a triple system leading to the current orbit \citep{2016MNRAS.456.3401P, 2021MNRAS.503.4276H}, supported by light echo observations \citep[e.g.,][]{2018MNRAS.480.1466S} and an extended central high-mass torus-like structure surrounding the central binary \citep{2017ApJ...842...79M}. In this scenario, the luminous blue variable primary star is currently orbited by a secondary star that is a classical Wolf-Rayet star, as discussed by \citet{2018MNRAS.480.1466S}. The system began as a hierarchical triple, and mass transfer led to the initial primary becoming a hydrogen-deficient Wolf-Rayet star. Mass transfer causes the orbits to become unstable, which leads to the merger and leaves behind the highly eccentric binary system we see today. {An alternate model for the eruption relies on the fact that $\eta$ Car is a binary in a highly eccentric orbit, and proposes that the periastron events triggered large mass transfer events that caused the eruptions \citep{2010ApJ...723..602K}. A similar model was used to explain the much less massive eruption that was seen from the SMC system HD 5980 during its LBV-like outburst \citep[e.g.,][]{2021A&A...653A.127K}.}

\begin{figure}
    \centering
    \includegraphics[angle=90,width=\columnwidth]{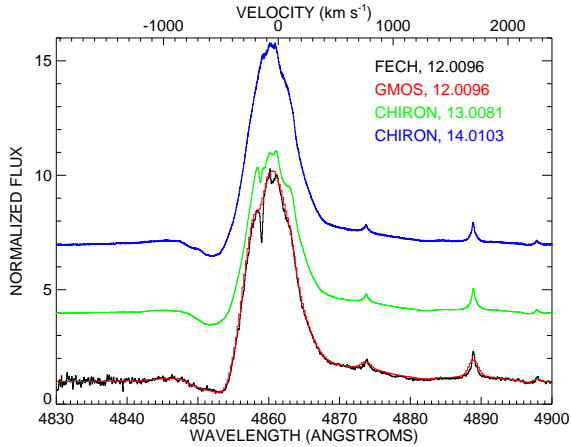}
    \caption{A comparison of an example Gemini-GMOS spectrum used by \citet{Grant2020} with the CTIO data from the fiber echelle (FECH) in 2009 and with more recent CHIRON data at the same phase (phases given in the legend). Note that the pixel sizes are indicated for the spectra, which is most obvious for the GMOS spectrum. The spectra are offset by orbital cycle, which highlights the complexities in the echelle spectra compared to the GMOS data.}
    \label{fig:GMOS}
\end{figure}

While the binary nature of the system was inferred by \citet{1996ApJ...460L..49D} and \citet{1997NewA....2..107D}, the orbit of the system has mostly eluded observers since the discovery of the spectroscopic events by \citet{Damineli1996}. \citet{1997NewA....2..387D} criticized the first orbit published by \citet{1997NewA....2..107D} and published a higher eccentricity model using the same data as \citet{1997NewA....2..107D}. Since these first attempts to derive the orbital motion of the system, very few observationally derived models have appeared in the literature, with most references to the orbit being inferred for modeling purposes. Recently, \citet{Grant2020} used archival moderate-resolution Gemini/GMOS spectra from 2009 to fit the hydrogen lines using multiple, weighted Gaussians to measure radial velocities corrected to account for motion from strong stellar winds. They derived a single-lined spectroscopic orbit {based on the upper Balmer lines} to be $T_0 = 2454848$ (HJD), $e = 0.91$, $K_1 = 69$ km s$^{-1}$, and $\omega_{\rm pri} = 241^\circ$ with the period of 2022.7 d that has been widely adopted based on multi-wavelength observations \citep[e.g.,][]{Teodoro2016}. These are broadly consistent with the smoothed-particle hydrodynamical (SPH) models used to describe variability across the electromagnetic spectrum \citep[e.g.,][]{2013MNRAS.436.3820M} including the X-ray light curves \citep[e.g.,][]{2008MNRAS.388L..39O}, optical He~{\sc{i}} absorption variability \citep{Richardson2016}, and the near-UV emission observed with the \textit{Hubble Space Telescope} \citep{2012ApJ...746L..18M}.

While the results of \citet{Grant2020} establish the orbital parameters with greater precision to date, there are potential issues with the determination of orbital elements from hydrogen lines in $\eta$ Car's spectrum, as the
strong wind of the primary causes the effective photospheric radius to be further out from the central star for lower energy transitions. Indeed, \citet{Grant2020} found better results with higher-order Balmer lines than with the optically thick H$\alpha$ or H$\beta$. This is a known effect for evolved Wolf-Rayet stars, where the observed semi-amplitude can change with the ionization potential of the line measured because lower-energy emission lines tend to form further out in the wind, where they are more likely to be perturbed by the companion star as seen in $\gamma^2$ Vel \citep{2017MNRAS.471.2715R}. This effect causes differences from the true orbital motion for lower energy transitions, making it difficult to determine accurate orbits \citep{Grant2020}. \citet{Grant2021} confirmed that their methods used for emission-line stars worked for the WR binaries WR 133 and WR 140 that have combined spectroscopic and interferometric orbits \citep{2021ApJ...908L...3R, 2021MNRAS.504.5221T}.

The primary star in the $\eta$ Car system is a luminous blue variable star, with the largest measured value for a mass-loss rate for a massive star with $\dot{M} = 8.5 \times 10^{-4} M_\odot {\rm yr}^{-1}$ and a terminal wind speed of $v_\infty = 420$ km s$^{-1}$ \citep{ 1997ARA&A..35....1D,2012MNRAS.423.1623G}.
Prior to the recent kinematic studies of \citet{Grant2020} and \citet{Grant2021}, the best constraints on the companion star parameters, while indirect, came from the X-ray variability analyses from {\it RXTE}, {\it Swift}, and {\it NICER} observations of the system \citep{2001ApJ...547.1034C, 2017ApJ...838...45C, 2022ApJ...933..136E}. These analyses point to a secondary star with a mass-loss rate on the order of $\dot{M}\sim10^{-5}M_\odot {\rm yr}^{-1}$ and a terminal velocity of $v_\infty \sim 3000$ km s$^{-1}$ \citep{2002A&A...383..636P}. These values are broadly in agreement with the suggestion based on the merger models and mass-loss parameters that the remaining secondary would be a Wolf-Rayet star. Despite recent work with long-baseline near-infrared interferometry by \citet{Weigelt2021}, no direct detection of the companion star has been made to date. From the interferometric data, a minimum primary-secondary flux ratio of $\sim$50 was derived {in the $K$-band \citep{2007A&A...464...87W}. Given the extreme luminosity of the LBV primary, this is consistent with any O or WR star in the Galaxy}.

The evolution of the secondary star may well have been significantly modified by interactions and mass exchange during formation of the present-day binary, but if the {current secondary star} is a classical H-free Wolf-Rayet star as suggested by \citet{2018MNRAS.480.1466S} and \citet{2021MNRAS.503.4276H}, or a hydrogen-rich WNh star, possibly the best line to detect it in the optical would be the He~{\sc{ii}} $\lambda$ 4686 line, which is the dominant line in the optical for the nitrogen-rich WR stars, or the hydrogen-rich WNh stars. Most of the observations of He~{\sc{ii}} were made near periastron, where the He~{\sc{ii}} excess can be explained by ionization of He~{\sc{i}} in the colliding winds in a highly eccentric binary. \cite{Teodoro2016} showed that the variability could be explained with the smoothed-particle hydrodynamics models of \citet{2013MNRAS.436.3820M}. Away from periastron ($0.04 < \phi < 0.96$), the He~{\sc{ii}} line is typically not observed with moderate resolving power and a nominal S/N of $\sim$100. 

In this paper, we present our analysis of the spectroscopy collected with the CTIO 1.5 m telescope and the CHIRON spectrograph, as well as the data collected with the previous spectrograph on that telescope with the aim of better constraining the kinematics of the system. These observations are described in Section 2. In Section 3, we review the variability in the two Balmer lines we can easily measure (H$\alpha$ and H$\beta$). Section 4 describes our techniques of measuring the radial velocity of the H$\beta$ line, and presents observations of He~{\sc{ii}} away from periastron in the hope of determining the orbit of the companion star. We discuss our findings in Section 6, and conclude this study in Section 7.
%Another even more elusive feature in the $\eta$ Carinae system is any detection of the companion star. The best parameters for the secondary have come from the X-ray analyses of REFERENCE(https://ui.adsabs.harvard.edu/abs/2001ApJ...547.1034C/abstract), who find that the wind parameters must be on the order of $\dot{M}\sim10^{-5}M_\odot {\rm yr}^{-1}$ and $v_\infty \sim 3000$ km s$^{-1}$. 

\section{Observations}

We collected high resolution spectra of $\eta$ Carinae during the periastron passages of 2009, 2014, and 2020. Many additional spectra were taken in the intermediate phases of the binary orbit as well. These were collected from the 1.5m telescope at Cerro Tololo Inter-American Observatory (CTIO 1.5) and both current CHIRON and the former fiber-fed echelle spectrograph (FECH). 
The data from the 2009 spectroscopic event spanned from 2008 October 16 to 2010 March 28, with approximately one spectrum taken every night between 2008 December 18 to 2009 February 19, which were previously used by \citet{Richardson2009, Richardson2015} {and cover the spectral range $\sim4700--7200$\AA.} These spectra with the fiber echelle\footnote{http://www.ctio.noao.edu/noao/content/CHIRON} were collected in late 2009 and 2010, and often had a signal-to-noise ratio around 80--100 per resolution element with $R \sim 40,000$. In total, we analyzed 406 spectra of the system.

The 2014--2020 data were collected with the new CHIRON spectrograph \citep{Tokovinin2013}, and spanned the time between 2012 March 2 and 2020 March 16, with high-cadence time-series spanning the 2014 and 2020 periastron passages between 2013 December 29 through 2015 April 21 as well as between 2020 January 3 to 2020 March 16 when the telescope shut down for the COVID-19 pandemic. {The CHIRON spectra cover the spectral range of $\sim$4500-8000\AA, with some spectral gaps between orders in the red portion of the spectrum.} The data covering the 2014 periastron passage were previously used by both \citet{Richardson2016} and \citet{Teodoro2016}. These data have a spectral resolution of $R\sim 80,000$ and typically have a signal-to-noise of 150--200 in the continuum and were all reduced with the CHIRON pipeline, which is most recently described by \citet{2021AJ....162..176P}. In addition to the pipeline reductions, we perform a blaze correction using fits from an A0V star, as done by \citet{Richardson2016}, allowing orders to be merged if needed. This process resulted in a flat continuum in regions that were line-free.

\begin{figure*}
	\includegraphics[width=\columnwidth]{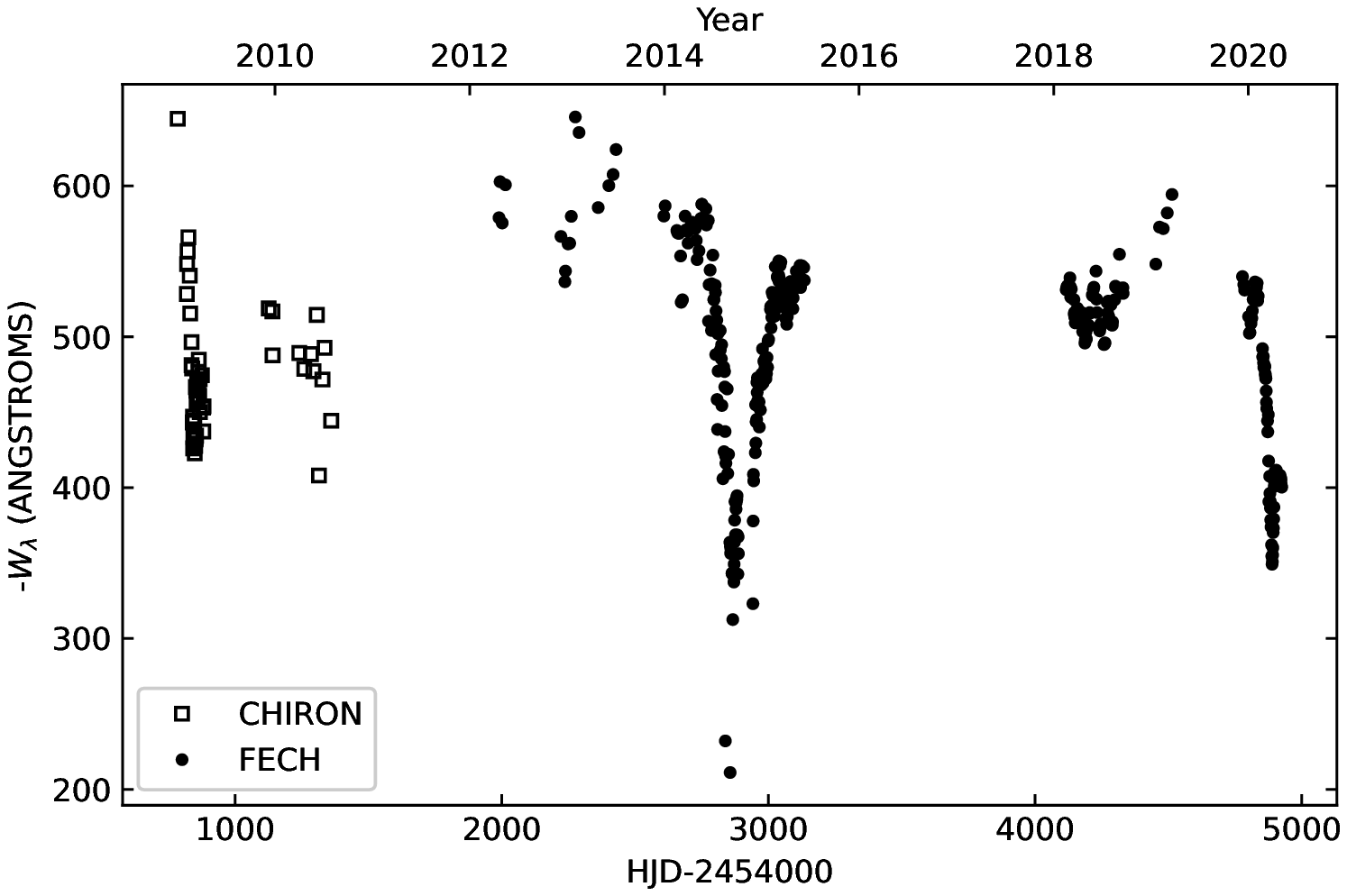}
	\includegraphics[width=\columnwidth]{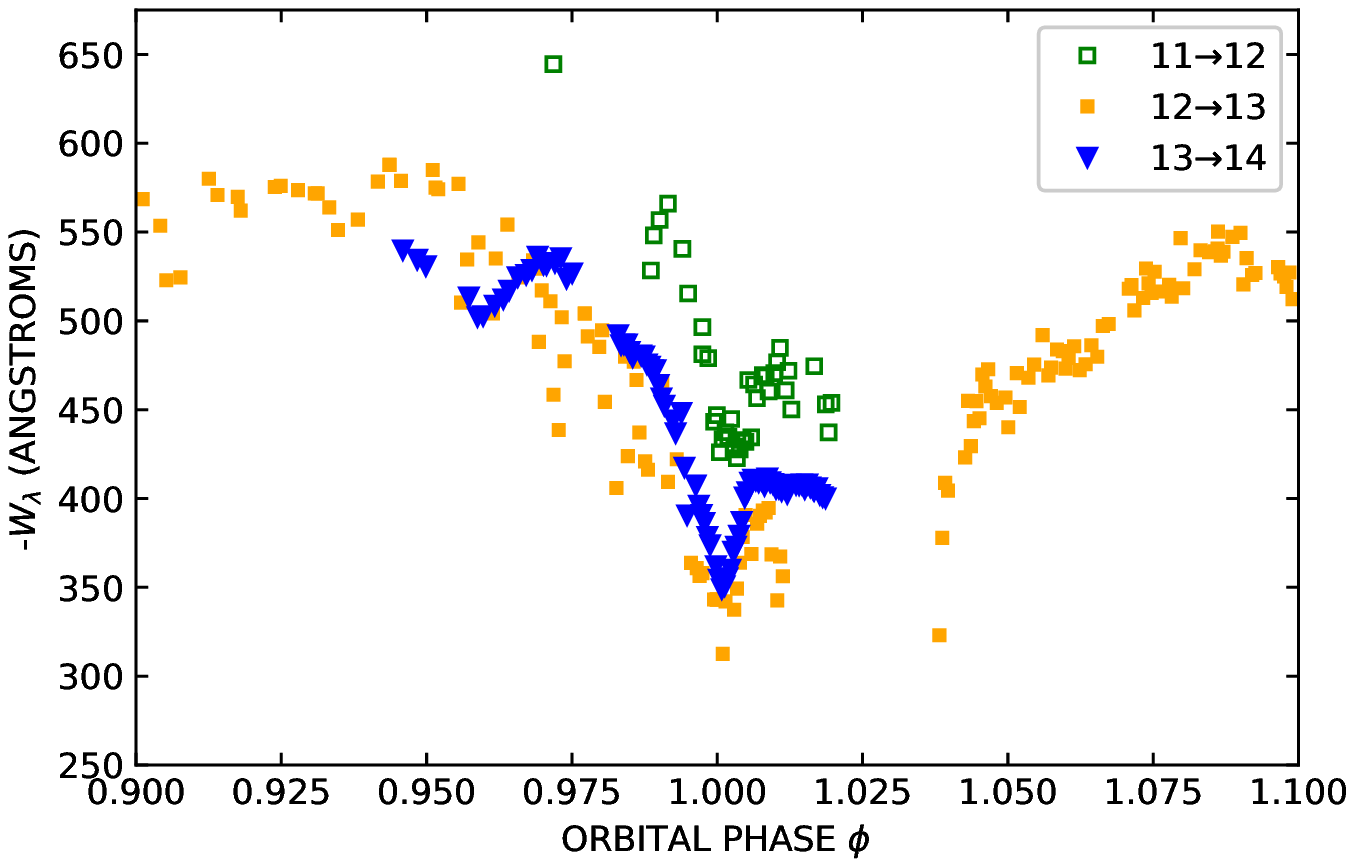}
    \caption{Variation in H$\alpha$ emission line with respect to time (left) and phase (right); with the data taken between October 2008 and March 2020. Data taken from the previous echelle spectrograph is indicated by open squares and data from the new CHIRON spectrograph is indicated by solid dots. In the phase plot, we show the different cycles in different colors to clarify the timing of each data set. Furthermore, the errors are typically the size of the points or smaller. {The phase convention shown in the right panel references the low-ionization spectrum near periastron first observed by \citet{1953ApJ...118..234G}.} }
    \label{fig:H alpha Long Time Series}
\end{figure*}

\begin{figure*}
	\includegraphics[width=\columnwidth]{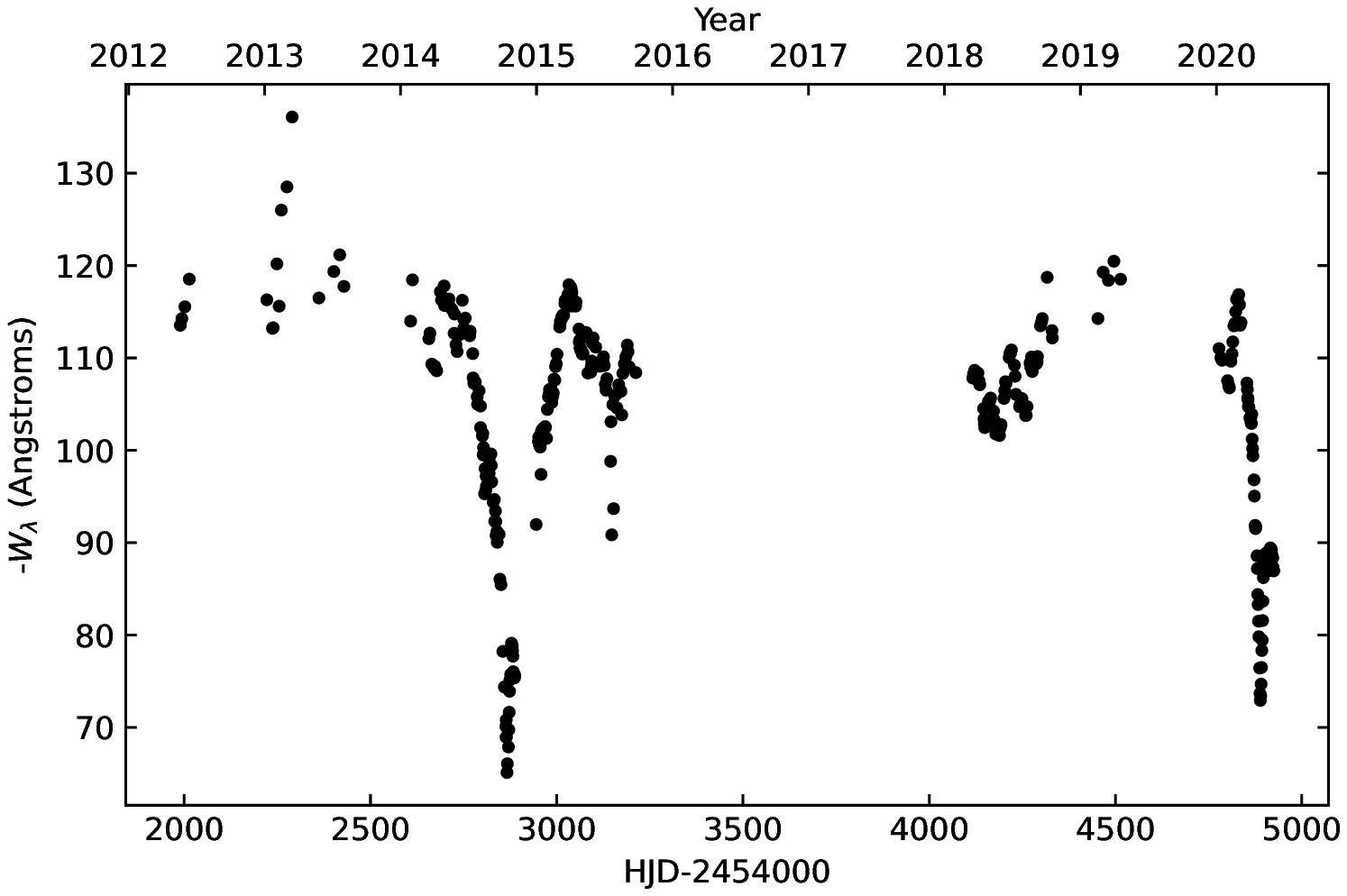}
	\includegraphics[width=\columnwidth]{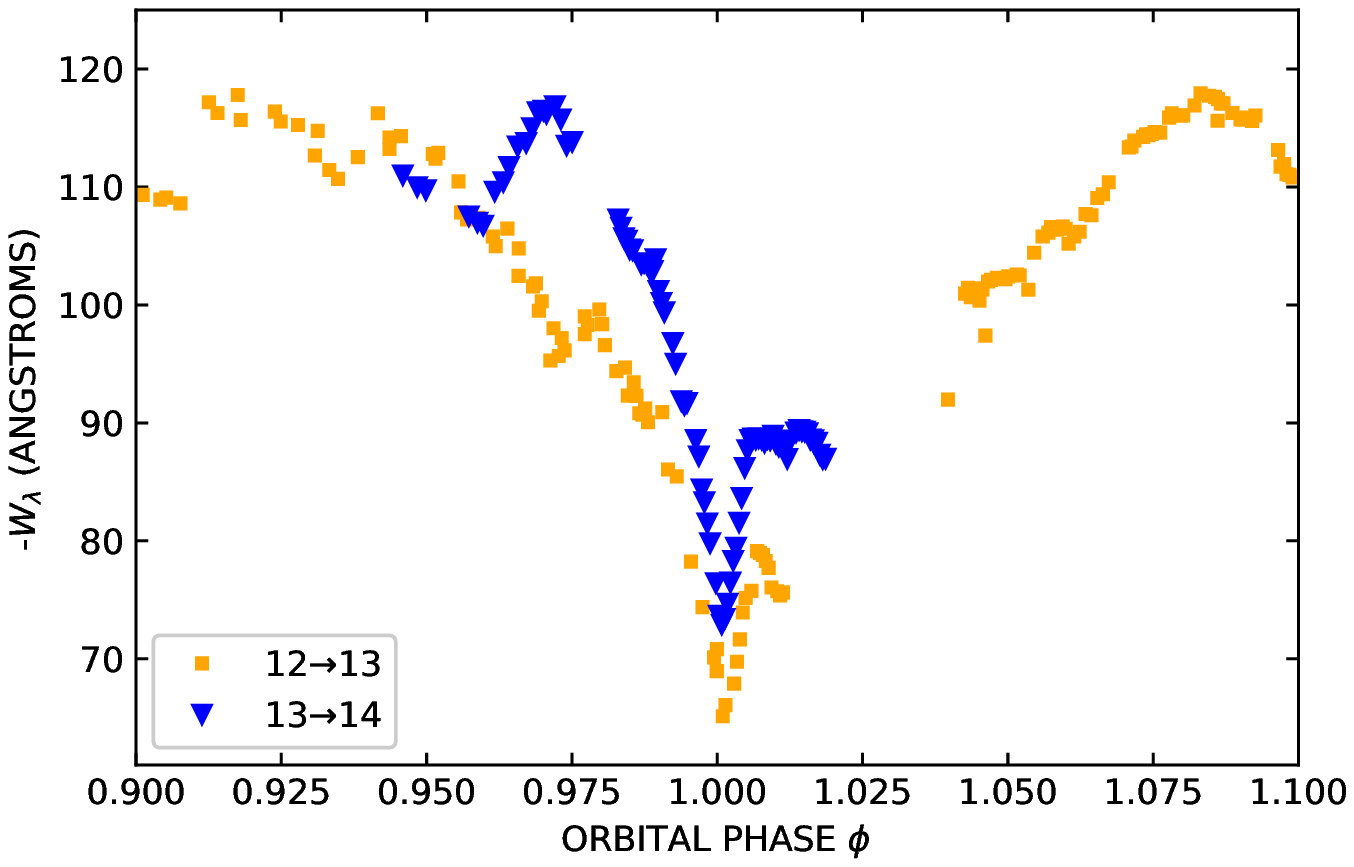}
    \caption{Variation in H$\beta$ emission line with respect to time (left) and phase (right); with the data taken with CHIRON spectrograph as the FECH data were too noisy to determine equivalent widths. In the phase plot, we show the two recent cycles in different colors to clarify the timing of each data set. Furthermore, the errors are typically the size of the points or smaller. {The phase convention shown in the right panel references the low-ionization spectrum near periastron first observed by \citet{1953ApJ...118..234G}.}}
    \label{fig:H beta Long Time Series}
\end{figure*}

These observations were all fiber-fed with the fiber spanning 2.7\arcsec\ on the sky, meaning that the data include the nebular emission from the Homunculus nebula formed from the eruption of $\eta$ Car in the mid-nineteenth century, as well as the Weigelt knots \citep{1986A&A...163L...5W} that are thought to have originated from the second eruption in the 1890s. The CHIRON spectra are normalized through a comparison with a measured blaze function from the star HR 4468 {(B9.5V)}, as was done in the analysis of \citet{Richardson2016}. Example spectra are shown in Figure.~\ref{fig:GMOS}, with a comparison to a spectrum used by \citet{Grant2020} and \citet{Grant2021}. 

\section{Measured variability in the Balmer lines, H$\alpha$ and H$\beta$}

Our observations are unique in providing both the spectral resolution and signal-to-noise to measure the line strength (equivalent width) and profile morphology of the emitting gas for the H$\alpha$ and H$\beta$ lines of $\eta$ Carinae. Here, we detail the observations of the variability of the hydrogen lines. We estimate errors on equivalent width using the methods of \citet{2006AN....327..862V}. {We note that the analysis of \citet{Richardson2015} includes many optical wind lines near the 2009 periastron passage and phases far from periastron. These line profiles all show minimum line strength near periastron as the secondary's high ionizing radiation goes behind the primary star's optically thick wind. We use a phase convention in which the low-ionization state observed by \citet{1953ApJ...118..234G} in 1948 is deemed to be cycle 1, so that the low-ionization state starting in Feb. 2020 marks the start of cycle 14.} {We leave the kinematics analysis of the metal lines for a future analysis in order to confirm the results of \citet{Grant2020} and \citet{Grant2021} here, with plans of using higher signal-to-noise spectra in a future analysis. }

\subsection{H$\alpha$}

\cite{Richardson2009} examined the variability of the H$\alpha$ profile of $\eta$ Carinae across the 2009 periastron passage. They found that the profile's strength decreased during the periastron passage and reached a minimum a few days following the X-ray minimum. They postulated that the changes were caused by the drop in the ionizing flux from the secondary when the companion moved to the far side. In addition, they observed an appearance of a P Cygni absorption profile and an absorption component at $-145$ km s$^{-1}$, that also appeared as the secondary's ionizing radiation was blocked by the primary star's optically thick wind. \cite{Richardson2015} expanded upon this model to describe the variations of the optical He~{\sc{i}} profiles while documenting the variability of the optical wind lines across the 2009 periastron passage.

We measured the equivalent width of H$\alpha$ for all of our spectra in the range 6500 -- 6650\AA. These results are shown in Fig.~\ref{fig:H alpha Long Time Series}, where we show the measurements both compared to time and to binary phase, assuming a period of 2022.7 d, and the epoch point given by \citet{Teodoro2016}, which represents the time of the periastron passage based on a comparison of the He~{\sc{ii}} observations \citep{Teodoro2016} to SPH models of the colliding winds. Broadly speaking, the strength of the line relative to the locally normalized continuum shows a fast decrease and recovery near each periastron passage. \citet{Richardson2009} found that the variability is smoother when considering the photometric flux in the determination of the equivalent widths. We did not make this correction in these data, but do see the similarities of the events in the context of the raw equivalent widths. 

There is no strong long-term variability in these observations, and the 2014 and 2020 observations were nearly identical in their variations. Recently, \citet{2019MNRAS.484.1325D, 2021MNRAS.505..963D} found that there are long-term brightness and spectral changes of the system that has been ongoing for decades and accelerated since the mid-1990s, but now seems to be stabilizing. The shape of the H$\alpha$ variability has remained similar over these three well-observed periastron passages, and the line strength has stabilized across the past two cycles, which could indicate that the system is mostly stable aside from the binary-induced variability. 

\citet{Richardson2009} also documented the timing of the appearance of the P Cygni absorption component for H$\alpha$. In the 2009 observations we see the absorption occurring at approximately HJD 2454840.7 ($\phi \approx 12.00$) and still persisting through the last observation, 2454881.7 ($\phi \approx 12.02$). In 2014 a P Cygni absorption occurs at 2456874.5 ($\phi \approx 13.00$) persisting until the object was not observable at HJD 2456887.5 ($\phi \approx 13.01 $). In 2020, the absorption is seen at 2458886.8 ($\phi \approx 14.01$) and still detected through the last observation on HJD 2458925.0 ($\phi \approx 14.02$).

A narrow absorption component was observed near $-145$ km s$^{-1}$ in the 2009 observations \citep{Richardson2009} from 2454836.7 ($\phi \approx 12.00$) through the last day of observation, 2454881.7 ($\phi \approx 12.02$). In 2014 an absorption in the same location is observed from 2456863.5 ($\phi \approx 13.00$) -- 2456977.8 ($\phi \approx 13.06$). There is no absorption at this location strong enough to make a definitive detection in 2020. \citet{2022arXiv220806389P} documented the changes in absorption behavior for the Na D complex at these velocities, showing that the absorption from these components associated with the Little Homunculus, formed during the second eruption in the 1890s, are weakening with time and moving to bluer velocities.

\begin{figure*}
	\includegraphics[width=7.5in]{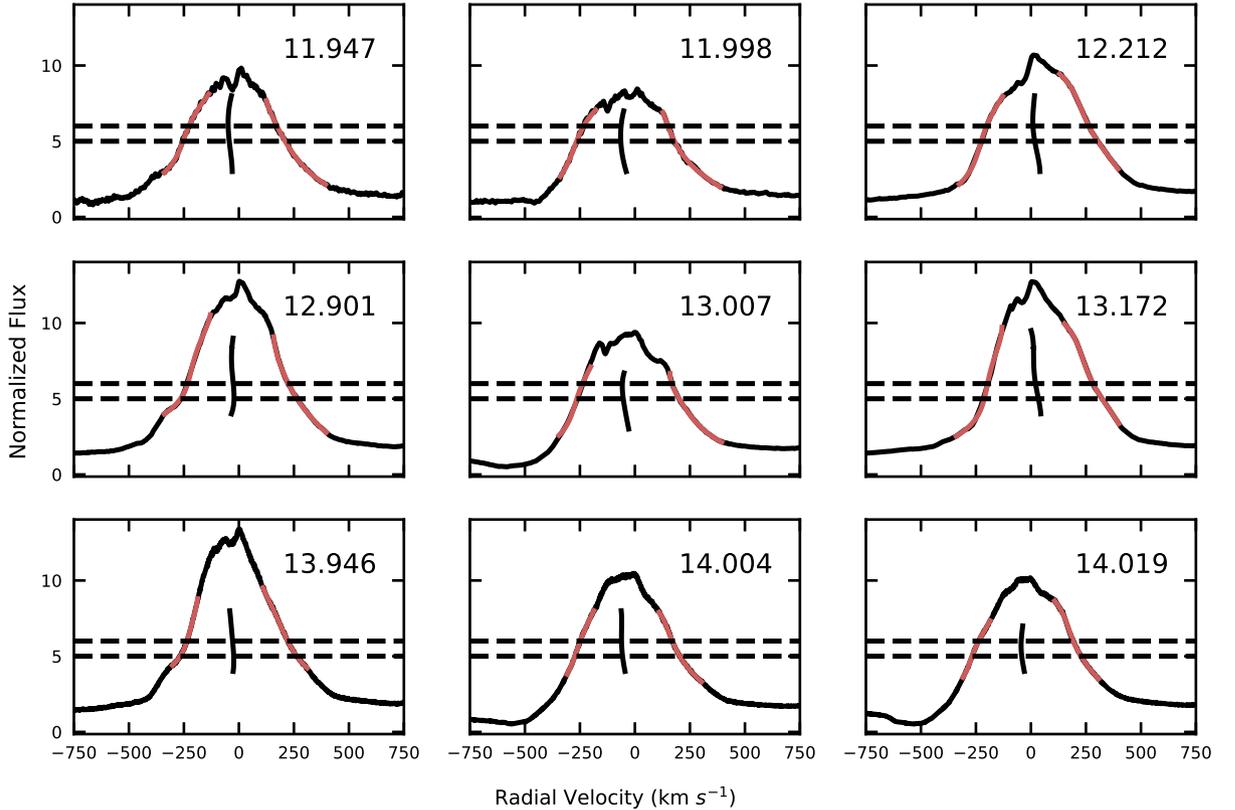}
    \caption{Example polynomial fits to H$\beta$ emission lines from 2009, 2014, and 2020 periastron events. The profiles are shown in black with the portion of the line wings fit with a polynomial shown in red. The bisector velocity is shown as a vertical line corresponding to the normalized flux at the same level as the measurements. Near the edges of these ranges, the bisector often appears to curve due to either profile asymmetries or larger errors in the polynomial fits. The bisector velocities between normalized flux levels of 5 and 6, indicated by the dashed lines, were averaged to obtain a final relative velocity for each day. Further details are given in Section 4.1.}
    \label{fig:H beta Bisect}
\end{figure*}

\subsection{H$\beta$}

While some of the H$\beta$ variability was documented for the 2009 periastron passage of $\eta$ Car by \citet{Richardson2015}, the full variability and timing of the changes is still not well documented in the literature. The lack of a more quantitative assessment of the variability is in part due to the lower signal-to-noise in the H$\beta$ data from the 2009 event. Similar to the H$\alpha$ profile, H$\beta$ experiences a P Cygni type absorption near $-500$ km s$^{-1}$ near periastron. We note the absorption appears in 2009 at approximately HJD 2454837.7 ($\phi \approx 12.00$) and persist through the last observation taken on 2454879.7 ($\phi \approx 12.01$). In 2014, it appears at approximately 2456863.6 ($\phi \approx 13.00$) and ends during a seasonal gap in observations beginning at 2456887.5 ($\phi \approx 13.01$). In 2020, the P Cygni absorption is observed between 2458886.8 ($\phi \approx 14.00$) and continues through the last day of observations on 2458925.0 ($\phi \approx 14.02$). This transient absorption was determined to be originating from the downstream bowshock by \citet{2022ApJ...933..175G}. 

A narrow absorption component, previously observed by \citet{Richardson2015}, is detected near $-145$ km s$^{-1}$ in the 2009 observations from 2454837.7 ($\phi \approx 12.00$) and proceeds through the end of observations on 2452879.7 ($\phi \approx 12.01$). In 2014, this absorption is observed between 2456864.5 ($\phi \approx 13.00$) and also persists through the last day of observations 2456887.5 ($\phi \approx 13.01$). As with H$\alpha$, there is no discernible absorption at $-145$ km s$^{-1}$ in 2020 observations.

Figure \ref{fig:H beta Long Time Series} shows the time series variation in the H$\beta$ equivalent width over the last two periastron cycles. We note that the 2009 observations are not included as they are recorded with the former echelle spectrograph and have lower signal-to-noise, though the appearance of the P Cygni absorption remains reliable. As with the H$\alpha$ equivalent widths, there is a consistency in the decrease in equivalent width for the time period corresponding to times close to periastron.

\section{Line Kinematics}
We measured the bisector velocity of H$\beta$ and the centroid position of the He~{\sc{ii}} $\lambda$4686 line. H$\beta$ measurements were taken during the 2009, 2014, and 2020 periastron events and the He~{\sc{ii}} 4686 measurements were taken for 2014 and 2018 and do not include time within $\phi$ = 0.95 -- 1.05 to avoid observations affected by periastron caused by colliding-wind effects which, to first order, behave with a $D^{-1}$ trend for adiabatic and $D^{-2}$ or steeper for radiative conditions, where $D$ is the orbital separation, which is small and quickly changing at periastron. \citet{Teodoro2016} show the behavior of the He~{\sc{ii}} 4686 line near periastron in detail. All measurements are tabulated in online supplementary data.

\subsection{Bisector velocities of H$\beta$}

The process used to find the bisector velocity of H$\beta$ is demonstrated in Fig.~\ref{fig:H beta Bisect}. \citet{Grant2020} and \citet{Grant2021} used a method of Gaussian decomposition using many components to moderate-resolution spectroscopy taken with Gemini-South and GMOS. Their GMOS spectra of $\eta$ Car are limited in that the highest resolving power available is $\sim 4400$, whereas our spectroscopy has a resolving power of $40,000$ from the fiber echelle, and $80,000$ for the CHIRON data. The profiles become more complex at higher spectral resolution, making this multiple-Gaussian method more difficult to implement, likely requiring more than twice as many components compared to the work of \citet{Grant2020}. 

\begin{figure*}
	\includegraphics[width=\columnwidth]{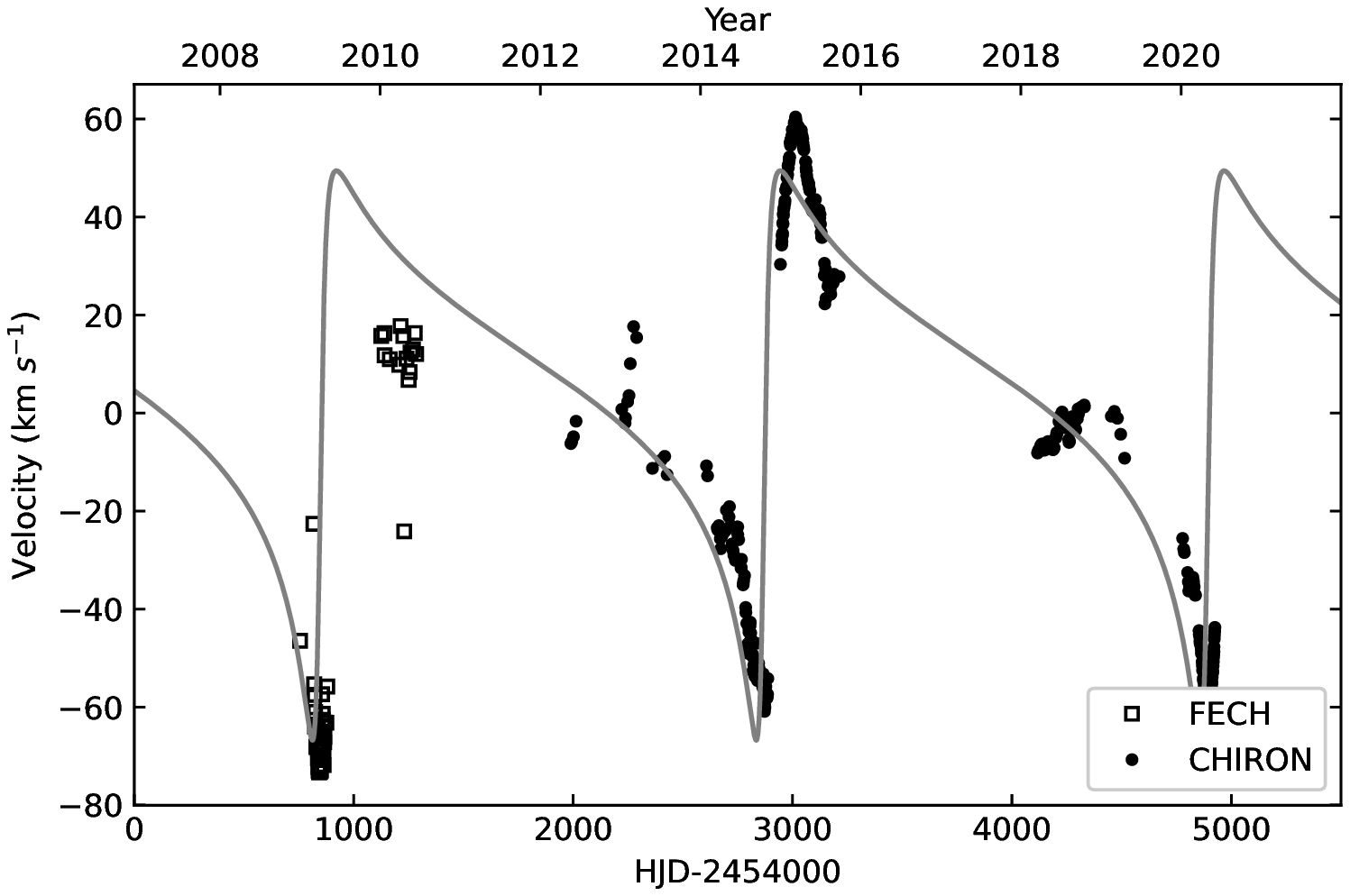}
	\includegraphics[width=\columnwidth]{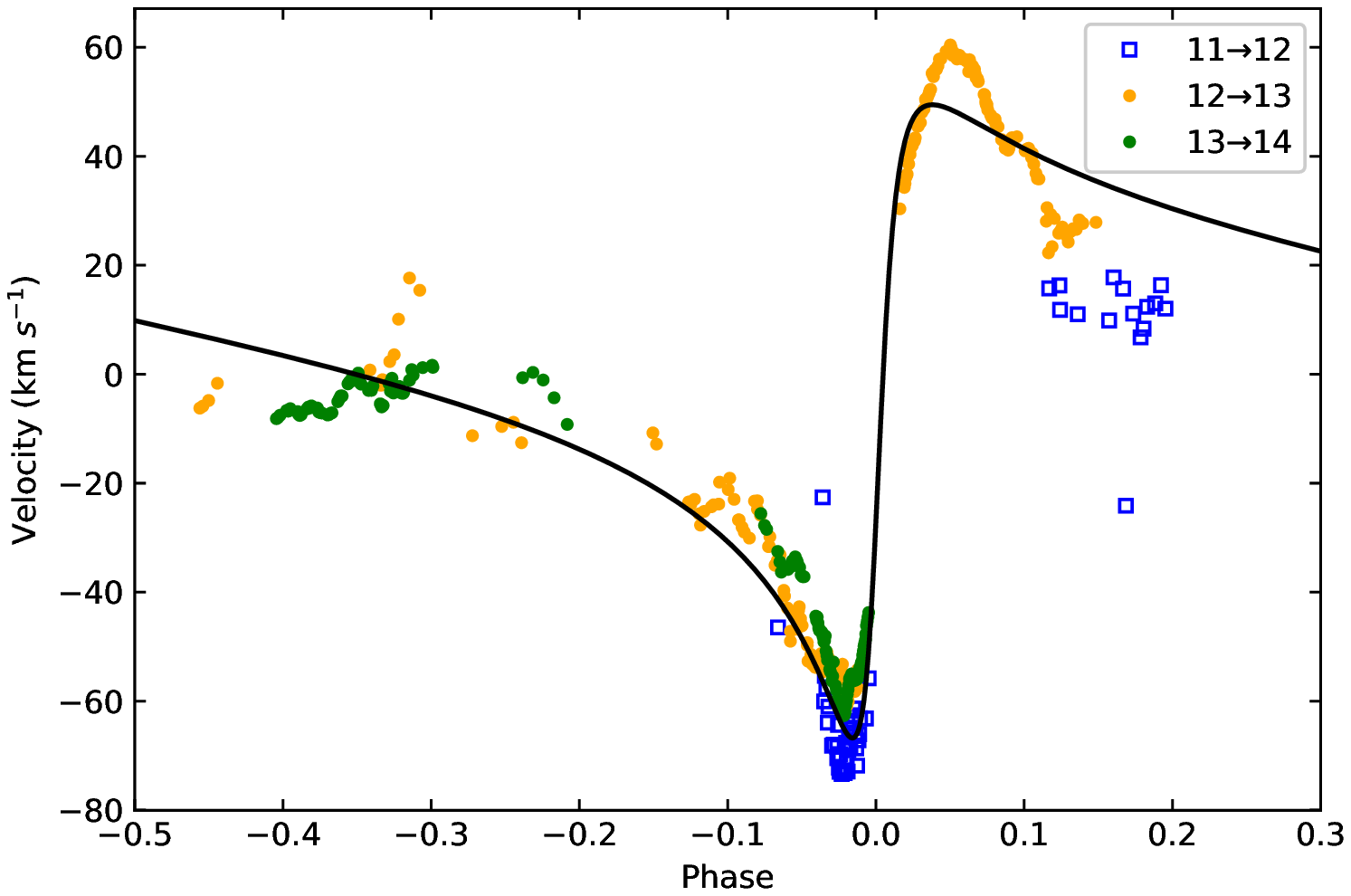}
    \caption{Radial velocities from H$\beta$ bisector measurements compared to time (left) and orbital phase (right). The orbital fit is described in Section 4.3 and typical errors are on the order of the size of the points.} 
    %during 2009, 2014 and 2020 periastron events.}
    \label{fig:H beta Velocity}
\end{figure*}

In order to create a simpler measurement that has reproducible results for any spectroscopic data set, we implemented a bisector technique. We began by fitting two fourth degree polynomials, {one to the red side and another on the blue side of the profile in order to smooth over any noise inherent in the data. Through this fit, we were then able to establish the bisecting velocity position at each emission level with higher precision. Example fits are shown in red in Fig.~\ref{fig:H beta Bisect}}. In the regions of heights of 4$\times$ the continuum up to 10$\times$ the continuum, we calculate the bisecting velocity. {This area was chosen based on the relatively vertical nature of the bisector in this region.} We then created comparisons of all spectra and found that the bisecting line was nearly always vertical in the region of $5-6\times$ the normalized continuum. We therefore used this region, measuring the velocity at every 0.1 increment between these values, and adopting an average measurement as the radial velocity for the spectrum. The choice of a common emission height with which to measure the bisector velocities allows us confidence in the results as it would relate to gas emitting from the same region for all spectra, whether the line is weak or strong in that particular observation. The resulting velocities are shown in Fig.~\ref{fig:H beta Velocity}. We provide this bisector code via GitHub\footnote{https://github.com/EmilysCode/Radial-Velocity-from-a-Polynomial-Fit-Bisector.git} for future use on comparable datasets. 

% We attempt implementing a simple yet effective method for determining radial velocity where we fit two polynomials to either side of the emission and then taking the bisector for every 0.1 interval between 4--10. These radial velocities are then able to be compared over time, when working with particularly large data sets a format such as a .gif is helpful in this process, to determine which region is the most appropriate representation of the radial velocity with least chance of possible manipulation from other effects. In these H$\beta$ measurements we determine the normalized flux region of 5 -- 6 to be the most appropriate representation of velocity and average the associated velocities, these resulting velocities are shown in figure 6.

% We use an error of ERROR, determined from the standard deviation. This value is appropriate to account for any keplerian motion caused by the winds.

% Figure 6 shows the long time series of the resulting radial velocities, and figure 7 shows a phase comparison of these velocities.

\subsection{He~{\sc{ii}} $\lambda$4686}

%%%%%%%%START HERE!!!%%%%%%%%%

\begin{figure}
	\includegraphics[width=\columnwidth]{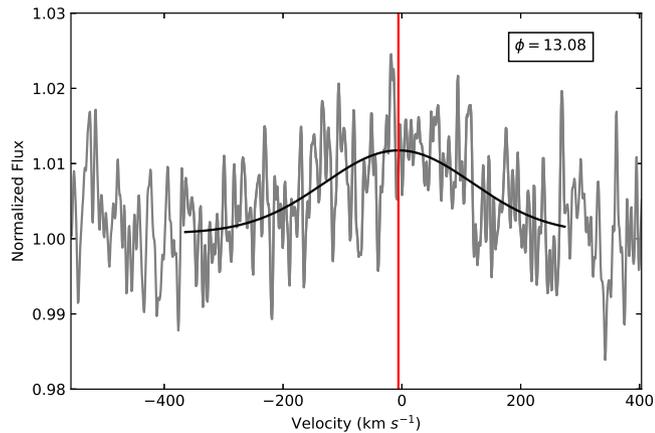}
    \caption{Gaussian fit to an example He~{\sc{ii}} emission line with a vertical line plotted at the fitted peak. This particular spectrum had a signal-to-noise ratio of 210 per resolution element. }
    \label{fig:HeII Fit}
\end{figure}

The region surrounding, but not blended with, the He~{\sc{ii}} $\lambda$4686 transition is complicated by several features including narrow emission lines from the Weigelt knots \citep{1986A&A...163L...5W} along with wind emission from Fe~{\sc{ii}} and He~{\sc{i}} lines \citep[for a figure showing that region of the spectrum, see][]{Teodoro2016}. While these do not directly overlap with the core of the He~{\sc{ii}} line, they can complicate this fitting if not properly avoided. The He~{\sc{ii}} $\lambda$4686 line has usually been observed near periastron passage when the line is dominated by the wind-wind collisions, which has been documented and modeled by \citet{Teodoro2016}. The line was discovered by \citet{2004ApJ...612L.133S}. Since then, multiple studies have attempted to explain the formation of the stronger line observed near periastron \citep[$L_{\rm He~{\sc{ii}}} \sim 300\  L_\odot$;][]{2006ApJ...640..474M,2011ApJ...740...80M,2015A&A...578A.122M,2012ApJ...746...73T,2015ApJ...801L..15D}, but the colliding wind model best reproduces the emission near periastron. This emission is strongest for times within $\pm 0.05$ in phase from periastron, as detailed in the recent analysis of \citet{Teodoro2016}. 

Outside of the phase intervals near periastron, the He~{\sc{ii}} $\lambda$4686 line could only be properly observed with high spectral resolution and high signal-to-noise data \citep{Teodoro2016}. Our data taken with CHIRON, after the 2014 periastron passage has the necessary sensitivity to detect this notably weak emission line. We measure the radial velocity of this line 
%We track the notably weak He~{\sc{ii}} 4686 emissions 
outside of $\phi = \pm 0.05$ of periastron, so that it minimizes the effects of the colliding winds that peak at periastron. %The He~{\sc{ii}} 4686 emission is too weak to be observed in the lower resolution data taken in 2009, and all observations made in 2020 were not within the phase range used. As such, only data taken in 2014 and 2018 was used.

\begin{figure}
	\includegraphics[width=2.4in,angle=90]{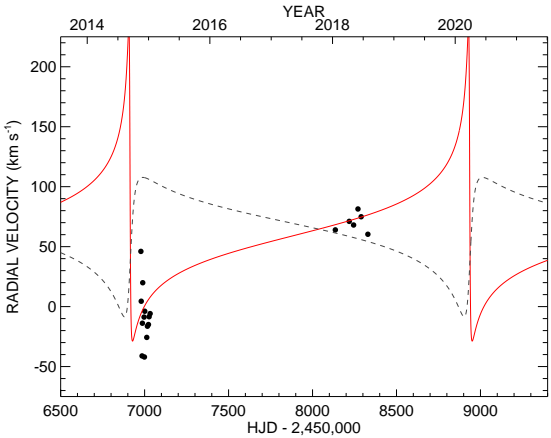}
    \caption{Radial velocity as determined using centroid positions in He~{\sc{ii}} emission at phases away from periastron during 2014--2018 with CHIRON. {We overplotted the He~{\sc{ii}} orbit from Table \ref{Orbital elements}, along with the H$\beta$ solution from our work shifted to the same $\gamma$-velocity as the He~{\sc{ii}} orbit as a grey dashed line. }}
    \label{fig:HeII Vel Time}
\end{figure}

As shown in Fig.~\ref{fig:HeII Fit}, we fit a Gaussian to the He~{\sc{ii}} emission line and use the centroid position to determine the radial velocity. Unfortunately, the continuum placement for the feature is not reliable enough to measure equivalent widths with precision, but the line was nearly constant in equivalent width when considering the errors of these measurements. Before fitting the 2018 observations near apastron, we needed to average up to ten observations to improve the signal-to-noise ratio. The resulting velocities are shown in Fig. \ref{fig:HeII Vel Time} with a resulting total of 19 data points. The averaging of the points from the 2018 data resulted in a smaller dispersion of the data than seen in the earlier points.

{The He~{\sc{ii}} line is normally absent in the spectra of luminous blue variables. The extreme mass-loss rate of $\eta$ Car does not preclude this emission line originating in the primary star's wind, as there are some combinations of parameters used that can create this weak emission feature in CMFGEN models. These models and parameters are very sensitive and depend on the mass-loss rate and stellar radii used. The He~{\sc{ii}} can be formed through strong wind collisions at times close to periastron \citep[e.g.,][]{Teodoro2016}. However, this line moves in opposition to the primary star's motion, so we consider this feature as originating from the companion during these phases far from periastron for the remainder of this analysis.}

\subsection{Orbital Kinematics and Observed Elements}

\begin{table*}
    \centering
\begin{tabular}{ |c c c c c c c| } 
 \hline
 Line & $T_0$ (HJD-2400000) & $e$ & $K$ (km s$^{-1}$) & $\omega$ (degrees) & $\gamma$ (km s$^{-1}$) & %v (km s$^{-1}$) & $a \sin i$ (R$_\odot$)& 
 Source \\
 \hline
 Pa$\gamma$ & 48800 $\pm$ 33 & 0.63 $\pm$ 0.08 & 53$\pm$6 & 286 $\pm{6}$ & $-15\pm 3$ & \citet{1997NewA....2..107D} \\
 
 Pa$\gamma$, He~{\sc{i}} 6678 & 48829$ \pm$ 8 & 0.802 $\pm$ 0.033 & 65.4 $\pm$ 3.5 & 286 $\pm$ 8 & -12.1 $\pm$ 2.7 & \citet{1997NewA....2..387D} \\
 
 Pa$\gamma$, Pa$\delta$ & 50861 & 0.75 & 50 & 275 & -12 & \citet{2000ApJ...528L.101D} \\
 
  H$\beta$ & 54854.9 $\substack{+4.5\\-4.1}$ & 0.82 $\pm{0.02}$  & 53.0 $\substack{+2.1\\-1.9}$  & 254 $\pm{4}$  & -25.5 $\pm{2.0}$  & \citet{Grant2020} \\
 
 All Balmer lines  & 54848.3 $\pm{0.4}$  & 0.91 $\pm{0.00}$  & 69.0 $\pm{0.9}$  & 241 $\pm{1}$  & $\ldots$ & \citet{Grant2020} \\
 
 Upper Balmer lines & 54848.4 $\pm{0.4}$  & 0.89 $\pm{0.00}$  & 69.9 $\pm{0.8}$  & 246 $\pm{1}$  &  $\ldots$ & \citet{Grant2020} \\

 H$\beta$ & 56912.2 $\pm{0.3}$ & 0.8100 $\pm{0.0007}$ & 58.13 $\pm{0.08}$& 251.43 $\pm{0.19}$& 6.34 $\pm{0.10}$ & %-& 1362 & 
 This work({\tt BinaryStarSolver})\\
 
 H$\beta$ & 56927.4 $\pm{0.5}$ & 0.8041 $\pm{0.0008}$  &  54.6$\pm$0.2 & 260.6 $\pm{0.2}$ & 4.83 $\pm 0.09$   & This work ({\tt PHOEBE})\\

 He~{\sc{ii}} & 56973.5 $\pm 0.2$  & 0.937 $\pm 0.001$  & 129.5$\pm$5.0 & 80.6 (fixed) & 63.1 $\pm 0.4$ &  This work \\
 \hline
\end{tabular}
\caption{Orbital elements from previous publications and the results from this work. For the orbits of \citet{Grant2020}, \citet{Grant2021}, and our work, the period has been held constant at 2022.7 d, while it was fit in the earlier work of \citet{1997NewA....2..107D}, \citet{1997NewA....2..387D}, and \citet{2000ApJ...528L.101D} with periods that agree with 2022.7 d within their errors. Note that our errors from the PHOEBE code may be underestimated, especially for the He~{\sc{ii}} line (see text for details). }
\label{Orbital elements}
\end{table*}

\begin{figure*}
    \centering
    \includegraphics[width=2.0\columnwidth]{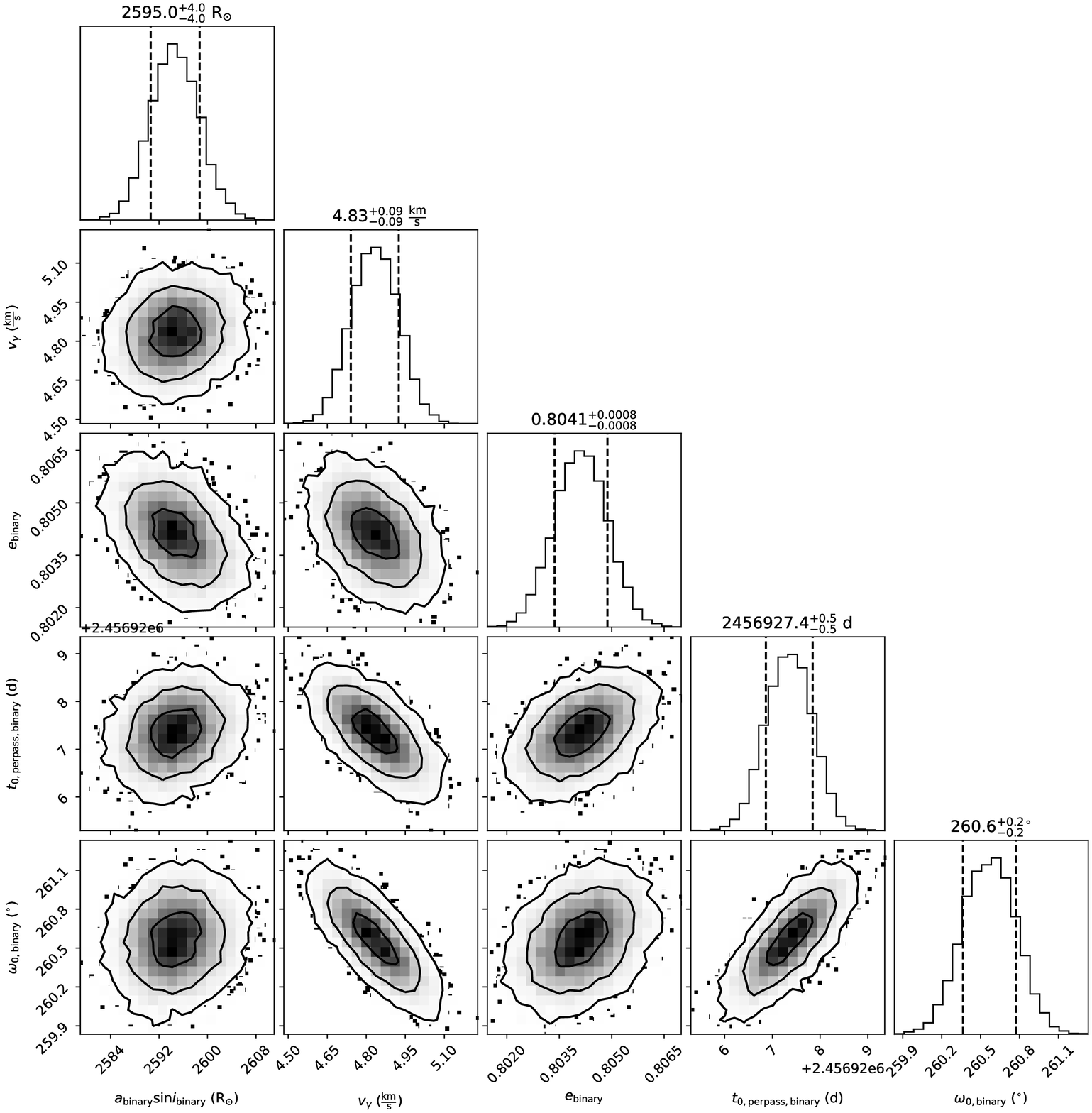}
    \caption{Results of the Markov chain Monte Carlo fit for the H$\beta$ velocities. Note that $\omega_0$ refers to the value of $\omega$ for the primary star. }
    \label{fig:mcmc}
\end{figure*}

We began our fit of the kinematics {of the primary star} with the {\tt BinaryStarSolver} software \citep{2020arXiv201113914M, 2020ascl.soft12004B}. The resulting orbit is broadly in agreement with the orbit derived with H$\beta$ velocities by \citet{Grant2020}, with the orbital elements given in Table \ref{Orbital elements}. Our resulting fits are in agreement with those of \citet{Grant2020} so we did not perform the same correction for the stellar wind effects as in their analysis.

In an attempt to fully assess the errors of the parameters, we used the {\tt PHOEBE} code \citep[PHysics Of Eclipsing BinariEs;][]{2005ApJ...628..426P, 2016ApJS..227...29P} to verify the orbital elements. The latest version of PHOEBE incorporates the Markov Chain Monte Carlo package {\tt emcee} \citep{2013PASP..125..306F}. Unlike traditional orbit fitting routines, PHOEBE fits using the variable of the projected semi-major axis ($a \sin i$) rather than the semi-amplitude $K$, but these are easily interchangeable using
$$a \sin i = \frac{(1-e^2)^{1/2}}{2\pi} K P.$$

These orbital elements are also similar to the other published orbital elements measured with H$\beta$, and the resulting orbit is shown in Fig.~\ref{fig:H beta Velocity}. The distribution of the errors from the Monte Carlo simulation, shown in Fig.~\ref{fig:mcmc}, is tightly constrained but shows that various orbital elements have errors that are interdependent with other parameters. While this represents the best solution to the entire data set, we explored how the parameters change if we kept only the densest of the three periastra observed (the 2014 event). Running the PHOEBE code with the MCMC package on just those data resulted in the eccentricity being slightly larger ($e=0.824$), the time of periastron being later (HJD 2,456,935.31), and the value of $a \sin i$ (hence $K_1$) being slightly larger at 2620.4 $R_\odot$. These values are outside the limits given with our MCMC fit of all of the data, so we caution that the errors in Table \ref{Orbital elements} are likely underestimated. We include the fit parameters in the same style as Fig.~\ref{fig:mcmc} in the online Fig.~\ref{fig:mcmc2014}.

Once the orbital elements for H$\beta$ were fit, we proceeded to run a simpler model for the He~{\sc{ii}} emission. For this PHOEBE model, we keep $\omega$ constant to that representing the primary star from the upper Balmer line results from \citet{Grant2020}. However, we do allow the semi-major axis, $\gamma$-velocity, $e$, and time of periastron passage to vary. The resulting orbit is more eccentric than that of the primary star when derived using H$\beta$ {(and a bit more eccentric than the \citet{Grant2020} solution)} and is shown in Fig.~\ref{fig:HeII Vel Time}. With future observations of the He~{\sc{ii}} line at times away from periastron, a combined double-lined orbit of the system with $\omega$ being consistent for the two stars will be possible.

\section{Discussion}

The optical spectrum of $\eta$ Car is dominated by emission lines from the wind of the primary and its ejecta. The dominant emission lines are the hydrogen Balmer lines, but there are strong lines from He~{\sc{i}} and Fe~{\sc{ii}} in the spectrum as well. The He~{\sc{i}} lines, when considered in non-LTE stellar wind models, are a strong
function of the adopted value of the stellar radius. However, if most of the
He~{\sc{i}} emission comes from the colliding wind interaction region, it forces a larger stellar core radius value for the primary star, $\sim 120 R_\odot$ in the preferred models \citep[see][for many further details]{2001ApJ...553..837H}.
The model of \citet{2012MNRAS.423.1623G} improved previous spherically symmetric models of \citet{2001ApJ...553..837H} in that the spectrum was modeled with a cavity carved from the wind of the secondary, {which was included} along with a central occulter or ``coronagraph" that extended $\sim0.033$\arcsec\ to allow for stronger He~{\sc{i}} emission, and better agreement for the P Cygni absorption lines. Given the spectral modeling agreement for the spectroscopically similar star HDE 316285 \citep{1998A&A...340..483H}, the strong disagreements for the absorption components and He~{\sc{i}} lines led to an interpretation that the He~{\sc{i}} lines are formed in the wind-wind collision region of the system \citep{2007ApJ...660..669N}. Indeed, the P Cygni absorption component variability of the optical He~{\sc{i}} lines seems to represent the outflowing shocked gas from the wind-wind collision region \citep{Richardson2016}. These results all indicate that the best lines in the optical for determination of the orbit may indeed be the {upper} hydrogen Balmer lines, even if they are likely modified by the wind collisions. 

%Binary motion in the $\eta$ Carinae system has been notoriously difficult to measure. The first indications of binarity came with the discovery of the 5.54-year period by \citet{1996ApJ...460L..49D} and \citet{1997NewA....2..107D}. The wind of the primary star is very optically thick, meaning that we can only attempt to measure stellar kinematics from lines formed in the outflow. The first determination of an orbit came from \citet{1997NewA....2..107D}, who measured radial velocities from the Pa-$\gamma$ and Pa-$\delta$ lines. The more recent analyses of \citet{Grant2020} and \citet{Grant2021} use the optical Balmer series for orbital determination. The analysis of \citet{Grant2020} relies on a weighted mean of many Gaussian emission (and occasionally absorption) components. 
All of the measured orbits, including ours, rely on measurements taken when the line profiles are most variable near periastron. This likely causes additional errors in the parameters derived, but we tried to always sample emission from the same line formation region by taking bisector velocities at the same height. Furthermore, our technique produces nearly the same orbital elements as those from \citet{Grant2020} {in the case of H$\beta$.  \citet{Grant2020} proceeded to correct the orbital elements by considering the effects of the outflowing wind.}

These results all show that the system is a long-period and highly eccentric binary where the primary star is in front of the secondary at periastron, causing the ionization in our line of sight to drop during the ``spectroscopic events" {due to a wind occultation of the secondary at these times}. The results of \citet{Grant2020} show that the higher-order Balmer lines give different results than that of the lower-level lines such as H$\alpha$ or H$\beta$, which is expected as the higher level lines form deeper in the wind \citep[e.g.][]{2001ApJ...553..837H}. As such, the results of \citet{Grant2020} and \citet{Grant2021} should be considered the best for the primary star at the current time. Similar differences in the orbital kinematics is sometimes inferred for Wolf-Rayet stars \citep[e.g., $\gamma^2$ Vel;][]{2017MNRAS.471.2715R}. 

Despite the detection of the He~{\sc{ii}} $\lambda$4686 emission at times near apastron by \citet{Teodoro2016}, the exact formation channel for this line remains unclear. The emission lines in colliding wind binaries often vary as a function of the orbit due to the colliding wind line excess \citep[e.g.,][]{2000MNRAS.318..402H}, and the modeling of these variations has been done in the context of the so-called L\"{u}hrs model \citep{1997PASP..109..504L}. Recently, the excess emission was observed to be a strong cooling contributor when X-ray cooling becomes less efficient in the colliding wind binary WR\,140 \citep{2021ApJ...923..191P}. In WR\,140, the L\"{u}hrs model was used by \citet{2011MNRAS.418....2F} to explain the variations in the C III $\lambda$5696 line near periastron. 

The L\"{u}hrs model can explain changes in the radial velocity and the width of the excess emission. As can be seen in Fig.~\ref{fig:HeII Fit}, {we detect the He~{\sc{ii}} line with} our spectra, but the actual characterization of this line will have large errors in line width due to the limited signal-to-noise for the detection in the spectroscopy. We used the models for WR\,140 \citep{2011MNRAS.418....2F} as a starting point, changing stellar and binary parameters as appropriate to the $\eta$ Carinae system to investigate if the He~{\sc{ii}} velocities in Fig.~\ref{fig:HeII Vel Time} were from colliding wind excess emission. For the velocity of the outflow, we can see that during the periastron passage of 2014, $\eta$ Car's outflow reached velocities faster than the primary star's wind speed based on the optical He~{\sc{i}} lines \citep{Richardson2016}, which are slower than the excess absorption seen to reach nearly 2000 km s$^{-1}$ in the meta-stable He~{\sc{i}} $\lambda$10830 line \citep{2010A&A...517A...9G}. With these velocities, we expect to see the observed amplitude of the excess increase between the times of 2015 and 2018 like we see in Fig.~\ref{fig:HeII Vel Time}, but with amplitudes of at least 1000 km s$^{-1}$, much greater than the $\sim100$ km s$^{-1}$ observed.

Therefore, the analysis of the He~{\sc{ii}} $\lambda$4686 emission line {at times away from periastron} from the CHIRON spectra is an important observation towards understanding the nature of the companion. We note that the data indicate a narrower emission line profile then expected from the parameters inferred for the secondary. However, the primary star dominates the spectrum, and the motion of this peak opposite the primary indicate that the He~{\sc{ii}} excess could be from the secondary's wind. {In particular, the L\"{u}hrs models of the kinematics of the He~{\sc{ii}} line seem to exclude the possibility that the line is formed in the colliding winds at times away from periastron.}

The models of \citet{2018MNRAS.480.1466S} suggest that the companion should be a classical Wolf-Rayet star. The classical hydrogen-free Wolf-Rayet stars can be split into the WN and WC subtypes. The WN stars show strong He and N lines, with the He~{\sc{ii}} $\lambda$4686 typically being the strongest optical line, whereas the WC subtype exhibits strong He, C, and O lines with the C IV $\lambda\lambda$5802,5812 doublet often being the strongest optical line. There is also the rare WO subtype, which is similar to the WC subtype but shows more dominant O lines. The WO stars were recently shown to have higher carbon and lower helium content than the WC stars, likely representing the final stages of the WR evolution \citep{2015A&A...581A.110T,2022arXiv220404258A}. Given the generalized characteristics of WR stars, a WN star would seem the most likely companion star if the He~{\sc{ii}} $\lambda$4686 line is from the companion at times further from periastron. 

For contrast, the Carina nebula is also the home to several hydrogen-rich Wolf-Rayet stars: WR 22, WR 24, and WR 25 \citep{2015MNRAS.447.2322R}\footnote{http://pacrowther.staff.shef.ac.uk/WRcat/}. This type of WR star tends to be considered the higher mass and luminosity extension of the main sequence. As such, these stars have masses in excess of $\sim 60 M_\odot$, with the R145 system in the LMC having masses of the two WNh stars being 105 and 95 $M_\odot$ \citep{2017A&A...598A..85S}. Like the classical WN stars, these stars have similar nitrogen and helium spectra, along with stronger emission blended on the Balmer lines which overlap with Pickering He~{\sc{ii}} lines. The region surrounding the He~{\sc{ii}} $\lambda$5411 line in our $\eta$ Carinae spectra does not exhibit emission lines at the same epochs as our observations of He~{\sc{ii}} 4686, making it difficult to quantify the companion's properties without the higher order He~{\sc{ii}} lines which would also be notably weaker than He~{\sc{ii}} $\lambda$4686.

With the assumption that the He~{\sc{ii}} orbit shown in Table \ref{Orbital elements} is from the companion star, and that the semi-amplitude from the higher-order Balmer lines for the primary star \citep{Grant2020}, then the semi-amplitude ratio shows that the primary star is 2--3 times more massive than the secondary star. This is also an indicator that the companion is not likely a WNh star, as that would imply the primary star could have a mass of in excess of 100 $M_\odot$. Models of the system, such as those by \citet{2008MNRAS.388L..39O} and \citet{2013MNRAS.436.3820M}, typically have the masses of the primary and secondary as 90 and 30 $M_\odot$ respectively, broadly in agreement with the kinematics of the orbits presented here. On the other hand, if $\eta$ Carinae A has a mass of $> 100 M_\odot$, the secondary would have a mass on the order of 50--60 $M_\odot$. This is similar to the nearby WNh star in the Carina nebula: WR22. The mass of this WNh star in an eclipsing system is 56--58 $M_\odot$ \citep{2022MNRAS.510..246L}.
The tidally-induced pulsations observed by \citet{2018MNRAS.475.5417R} were modeled with stars of masses 100 and 30 $M_\odot$, {and therefore may also support the higher masses suggested here}.

Most models for $\eta$ Car have a preferred orbital inclination of 130--145$^\circ$ \citep{2012MNRAS.420.2064M}, which agrees with forbidden [Fe~{\sc{iii}}] emission observed with {\it Hubble Space Telescope}'s Space Telescope Imaging Spectrograph. This inclination can be used with the mass function derived from the primary star's orbit 
$$f(M) = {{m_2^3 \sin^3 i}\over{(m_1+m_2)^2}} = (1.0361 \times 10^{-7})(1-e^2)^{3/2}K_1^3 P [{\rm M}_\odot]$$
to constrain the system's masses {with the mass function using the standard units measured and our measured H$\beta$ orbit using PHOEBE (Table 1).} The mass function is $f(M) = 8.30 \pm 0.05\ {\rm M}_\odot$, and would indicate a companion star with a mass of at least 60 $M_\odot$ if we assume a primary mass of $\sim$ 90 $M_\odot$. {Given the actual mass functions for the measured upper Balmer lines and He~{\sc{ii}} orbits,} the minimum masses required for these measured orbits are $M \sin^3 i = 102 M_\odot$ for the LBV primary and $M \sin^3 i = 55 M_\odot$ for the secondary, making the companion star's identification as a WNh star more likely.  These results are still preliminary and require follow-up observations to constrain the orbits. 

A WNh star can account for the mass of the secondary star in $\eta$ Car, but could cause some difficulty for the modeling of the Great Eruption models of \citet{2021MNRAS.503.4276H}. In that scenario, the companion star would be a hydrogen-stripped star, contrary to the hydrogen content of the WNh stars. Recently modeled WNh systems such as R144 \citep{2021A&A...650A.147S} show that the surface fraction of hydrogen is about 0.4. This does show some amount of lost hydrogen on the surface, so the scenario could still be relevant even if the final star is not a fully stripped classical Wolf-Rayet star, assuming that the evolution of the secondary star has not been significantly influenced by mass exchange prior to or during the merger event hypothesized by both \citet{2016MNRAS.456.3401P} and \citet{2021MNRAS.503.4276H}.

\section{Conclusions}

In this paper, we provide an orbital ephemeris for $\eta$ Carinae measured with a bisector method and high resolution ground-based spectroscopy of the H$\beta$ emission line, along with an ephemeris for the He~{\sc{ii}} $\lambda$4686 emission line at times far from periastron. Our findings can be be summarized as follows:

\begin{itemize}
    \item The H$\beta$ emission profile tracks the primary star, and our bisector method provides similar results as the multiple-Gaussian fitting method used by \citet{Grant2020}. The results show a high eccentricity orbit of the system with the primary star in front of the secondary at periastron.
    \item The weak He~{\sc{ii}} $\lambda$4686 emission tracks opposite the kinematics of the primary star, suggesting it is formed in the secondary star's wind at times away from periastron. This could support the hypothesis of the scenarios presented by \citet{2021MNRAS.503.4276H} for a stellar merger being the cause of the Great Eruption as the secondary could be a Wolf-Rayet star that has leftover hydrogen on its surface.
    \item With the assumed inclination of 130--145$^\circ$, the masses of the stars could be around $\sim$100 $M_\odot$ for the primary and at least 60 $M_\odot$ for the secondary. However, the mass ratio derived by comparing the two semi-amplitudes is about 1.9. New observations will be needed to better determine precise masses.
\end{itemize}

Future studies will be able to better measure the He~{\sc{ii}} 4686 orbit and refine its parameters. As shown in \citet{Grant2020}, the upper Balmer lines are more likely to reflect the orbital motion of the stars, and the upper Paschen lines will also be useful. However, our work shows that a simpler bisector measurement of higher resolution spectroscopy results in the same derived orbital elements as that of \citet{Grant2020}. Furthermore, with better signal-to-noise spectra, we can better determine if the He~{\sc{ii}} emission near periastron can be reproduced with a L\"{u}hrs model or if it is a signature of the companion. With this information, we will be able to more precisely measure the kinematics of the two stars and the  mass function, and then we can begin to better understand the current evolutionary status of the system. 

\section*{Acknowledgements}

We thank our referee, Tomer Shenar for many suggestions that improved this paper. These results are the result of many allocations of telescope time for the CTIO 1.5-m telescope and echelle spectrographs. We thank internal SMARTS allocations at Georgia State University, as well as NOIR Lab (formerly NOAO) allocations of NOAO-09B-153, NOAO-12A-216, NOAO-12B-194, NOAO-13B-328, NOAO-15A-0109, NOAO-18A-0295, NOAO-19B-204, NOIRLab-20A-0054, and NOIRLab-21B-0334. This research has used data from the CTIO/SMARTS 1.5m telescope, which is operated as part of the SMARTS Consortium by RECONS (www.recons.org) members Todd Henry, Hodari James, Wei-Chun Jao, and Leonardo Paredes. At the telescope, observations were carried out by Roberto Aviles and Rodrigo Hinojosa.
C.S.P. and A.L. were partially supported by the Embry-Riddle Aeronautical University Undergraduate Research Institute. E.S. acknowledges support from the Arizona Space Grant program. N.D.R., C.S.P., A.L., E.S., and T.R.G. acknowledge support from the {\it HST} GO Programs \#15611 and \#15992. AD thanks to FAPESP (2011/51680-6 and 2019/02029-2) for support. AFJM is grateful for financial aid from NSERC (Canada). The material is based upon work supported by NASA under award number 80GSFC21M0002. The work of ANC is supported by NOIRLab, which is managed by the Association of Universities for Research in Astronomy (AURA) under a cooperative agreement with the National Science Foundation.

%%%%%%%%%%%%%%%%%%%%%%%%%%%%%%%%%%%%%%%%%%%%%%%%%%
\section*{Data Availability}

All measurements can be found in Appendix~\ref{sec:appendix}. Reasonable requests to use the reduced spectra will be granted by the corresponding author.



\bibliographystyle{mnras}
\bibliography{References} % if your bibtex file is called example.bib

%\begin{thebibliography}{99}
%\end{thebibliography}

%%%%%%%%%%%%%%%%%%%%%%%%%%%%%%%%%%%%%%%%%%%%%%%%%%

%%%%%%%%%%%%%%%%% APPENDICES %%%%%%%%%%%%%%%%%%%%%

\appendix

\section{Appendix}
\label{sec:appendix}

\begin{figure*}
    \centering
    \includegraphics[width=2.0\columnwidth]{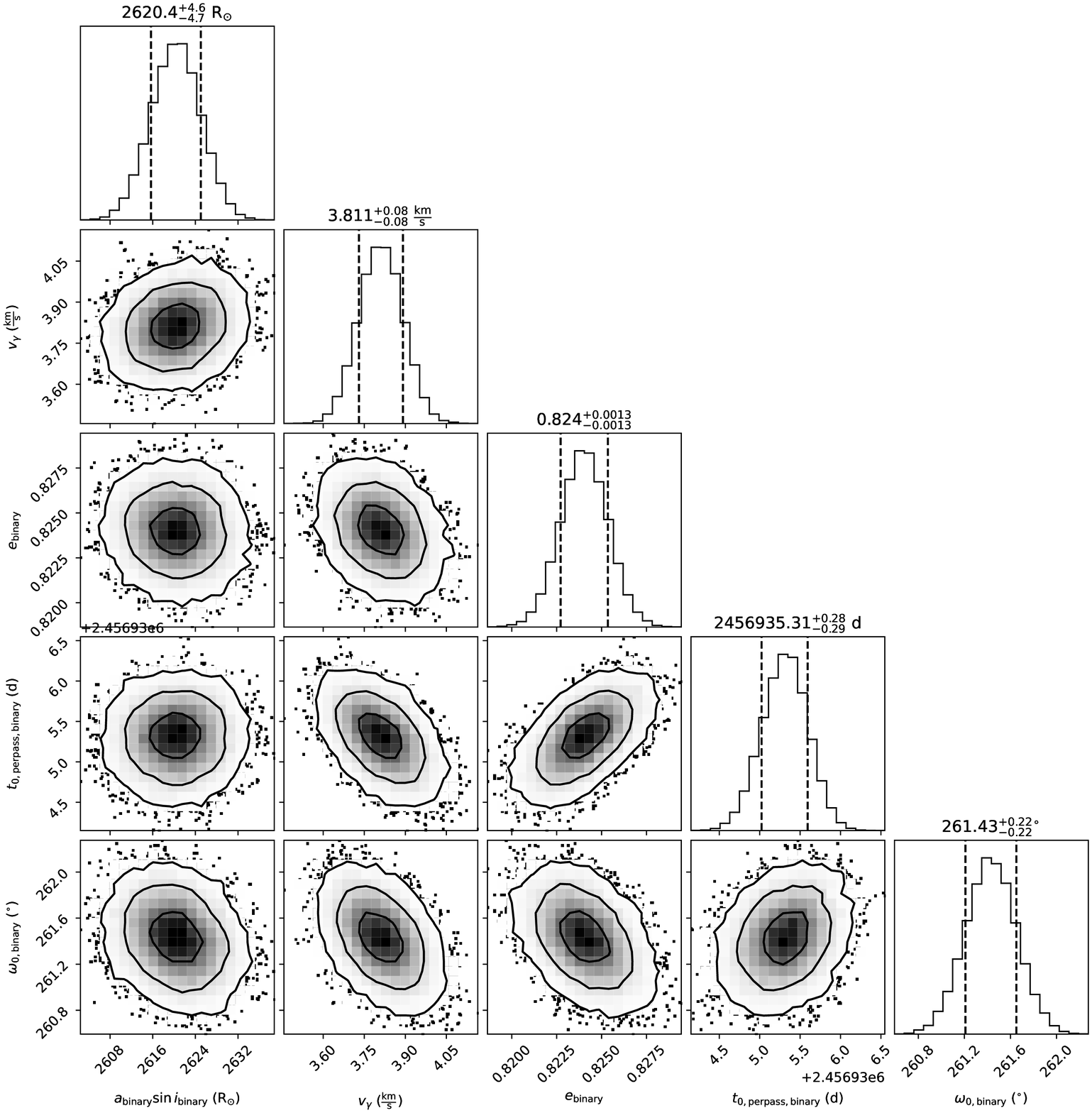}
    \caption{Results of the Markov chain Monte Carlo fit for the H$\beta$ velocities for only the data surrounding the 2014 periastron passage. Note that $\omega_0$ refers to the value of $\omega$ for the primary star. }
    \label{fig:mcmc2014}
\end{figure*}

\bsp	% typesetting comment
\label{lastpage}
\end{document}